\documentstyle[12pt,epsfig]{article}

\newcommand{\be}{\begin{equation}}
\newcommand{\ee}{\end{equation}}
\newcommand{\bea}{\begin{eqnarray}}
\newcommand{\eea}{\end{eqnarray}}

\begin{document}
\pagestyle{empty} 
\begin{flushright}
CERN-TH/97-370 
\end{flushright}
\centerline{\large \bf On the Index Theorem for Wilson Fermions}
\vskip 1.5cm
\centerline{\bf{P. Hern\'andez}}
\vskip 0.3cm
\centerline{ Theory Division, CERN, 1211 Geneva 23, Switzerland.}

\date{}
\abstract{ 
We consider a Wilson-Dirac lattice 
operator with improved chiral properties. We show
that, for {\it arbitrarily} rough gauge fields, it satisfies the index
theorem if we identify the zero modes with 
the small real eigenvalues of the fermion operator 
and use the geometrical definition of topological charge. This is also
confirmed in a numerical study of the quenched Schwinger model. These results
suggest that integer definitions of the topological charge based on counting  
real modes of the Wilson operator are equivalent  
to the geometrical definition. The 
problem of exceptional configurations and the sign 
problem in simulations with an odd number of dynamical Wilson fermions are 
briefly discussed.}
\vskip 3.cm
\begin{flushleft}
CERN-TH/97-370 \\
December 1997
\end{flushleft}
\vskip 0.5 cm
\vfill\eject
\pagestyle{empty}\clearpage
\setcounter{page}{1}
\pagestyle{plain}

\newpage
\pagestyle{plain} \setcounter{page}{1}

\section{Introduction}

The connection between gauge field topology and fermion zero modes is 
expected to have important physical consequences on the non-perturbative 
dynamics of baryon number violation in the SM or the breaking of the singlet 
chiral symmetry in QCD. The study of these effects, however, requires 
non-perturbative techniques and one would expect that  Monte Carlo methods on the 
lattice would ultimately be best suited to it. Unfortunately there is no proof of the 
Atiyah-Singer index theorem on the lattice, in spite of which, a big
effort has been devoted to the measurement of the topological susceptibility 
\cite{sus}. The main motivation for this measurement in $SU(3)$ 
is the relation of this quantity to the $\eta'$ mass, through 
the continuum 
formula of Witten and Veneziano \cite{wv}. This formula relies
on the validity of the index theorem, so it is important to understand 
to what extent this theorem survives on the lattice, where the connection 
between topology and fermion zero modes is not clear. 

It has been known for some time that there are remnants of the 
index theorem on the lattice for Wilson fermions. 
In \cite{smit} it was observed that in {\it smooth} gauge 
configurations, in which the geometrical definition
of topological charge \cite{geo} takes an integer value $Q^{geo}$, the Wilson-Dirac 
operator has $Q^{geo}$ small and {\it exactly} real eigenvalues with the appropiate chirality. The small real eigenvalues seem to 
play the role of the continuum zero modes. We will refer to the connection between the small real
eigenvalues of the Wilson-Dirac operator and the geometrical definition
of topological charge as the ``lattice index theorem'' (LITh). 
Unfortunately, when rough gauge fields are considered this connection seems to be lost. 

It is well known that the geometrical definition of topological charge
in four-dimensional Yang-Mills theories has dislocations. They are small 
, $O(a)$, objects, which carry topological charge, but have such a 
small plaquette action that they destroy the proper scaling 
of topological quantities constructed out of the geometrical charge.  
It has been argued that the geometrical definition of topological charge, 
being constructed
as an integral of a local density, is more sensitive to fluctuations of the
order of the lattice spacing than the number of fermion ``zero'' modes, 
which are non-local \cite{nara}. For this reason, there have been several 
proposals \cite{smit}\cite{nara} to measure topology by measuring the 
chiral charge of the small eigenvalues of the Wilson-Dirac operator.
The hope is that these definitions are free of dislocations and 
topological quantities show the expected scaling. 
If this were the case, then it is necessary that the LITh is violated, 
even in an average 
sense, as we approach the continuum limit, because the fermionic charge should 
scale differently than the geometrical one. 
The experimental fact is that the LITh is violated for rough gauge fields, but
satisfied for smooth ones. This does not allow us to draw any conclusion a priori 
on  whether the LITh will be satisfied in the continuum limit (in an average
sense), because rough fields are important even in this limit, since they 
are responsible for renormalization. 

In this letter, we present a  study of the LITh for a new fermionic action that was originally 
proposed to deal with chiral gauge theories \cite{us1}, which has
improved chiral properties at any fixed lattice spacing. We apply it here to 
the study of the Schwinger model. The new action is
constructed by interpolating the gauge variables smoothly, gauge invariantly,
and locally to a finer lattice \cite{us2}, in which the fermion determinant 
and propagator 
are defined in terms of the standard Wilson-Dirac operator. It can be shown 
that the factor of fine-graining controls the violations of chirality 
\cite{us1} like a power, to all orders in the gauge coupling
 (in the appendix, we present 
an explicit one-loop calculation in a $U(1)$ model in four dimensions of the additive
 renormalization to the fermion mass, which shows the expected suppression.). 
 The improvement in the validity of LITh is however much more dramatic than 
this power suppression might indicate. We find that, with a simple 
factor of fine-graining of $1/2$, the violations 
of LITh are absent for {\it arbitrarily} large couplings. This strongly 
indicates
that measuring topological charge by looking at the real eigenvalues
of the Wilson operator (when it has been properly improved) 
is equivalent to the geometrical definition.

If this is confirmed in four dimensions, the problem of dislocations should
 be handled by improving the 
plaquette action \cite{marga,hasen,us3} rather than by using a different 
{\it integer} definition of the topological charge. Any such definition 
is not sensitive to topological objects of size roughly of
 $O(a)$, since
$a$ is the only cutoff scale in the problem. Different definitions might 
have a slightly different cutoff, but generically there is no reason 
to expect that some of them will be affected by dislocations 
and not others, as long as their cutoff is of the same order \footnote{The 
fermionic charge defined in \cite{smit} is not an integer so this argument
does not apply to it.}.
The improved gauge action described in \cite{us3} is the natural choice to 
 combine with the improved fermionic action described here in order to 
ensure the proper scaling of the topological susceptibility constructed out
 of the geometrical or fermionic charges. 

In section 2, we review some known facts about the connection between 
zero modes in the continuum and real modes of the Wilson-Dirac operator 
and show that a lattice index theorem can be defined, which holds 
for every gauge configuration if we use the improved Wilson-Dirac operator. 
In section 3, we present results on the LITh for the standard action. 
In section 4, the improved action is described and the results on the LITh
are presented.
The small real eigenvalues of the Wilson-Dirac operator are related to the 
so-called exceptional configurations and are also responsible for the 
sign flips in the determinant of this operator, which make dynamical 
simulations with an odd number of fermions problematic. In section 5, we 
discuss the relevance of our results to these problems and conclude.

\section{Lattice Index Theorem}

We first review the known arguments that suggest the identification of
 the zero modes of the continuum Dirac operator with the 
exactly real eigenvalues of the Wilson-Dirac operator on the lattice \cite{smit,itoh,bardeen,gatt,gatt2}.

Continuum zero modes have a well-defined chirality. 
As was shown in \cite{smit,itoh}, the eigenmodes of the lattice
Wilson-Dirac operator with a non-vanishing chirality are necessarily real. This
is easy to see by realizing that the Wilson-Dirac operator satisfies, 
\bea
\gamma_5 \not\!\!D \gamma_5 = \not\!\!D^\dagger,
\label{gamma5}
\eea
which implies that eigenvalues come in complex conjugate pairs, i.e.
if $v_i$ is an eigenvector on the right of $\not\!\!D$ with eigenvalue $\lambda_i$, then $v^\dagger_i \gamma_5$ is an eigenvector on the left  
with eigenvalue $\lambda^*_i$. Then, it follows that, if $v^\dagger_i \gamma_5 v_i \neq 0$, 
\bea
\lambda_i = \lambda^*_i
\eea
must be satisfied and, vice versa, if $\lambda_i$ is complex then $v^\dagger_i \gamma_5 v_i = 0$. 
Generically, for real eigenvalues $v^\dagger_i \gamma_5 v_i = O(1)$. In fact
as found in \cite{smit}, this value is very close to $\pm 1$ for smooth 
backgrounds.

Another indication that real eigenvalues might have a topological origin is
their stability under perturbations. 
Consider the Wilson-Dirac operator $\not\!\!D = \not\!\!D^{(0)} + \epsilon \not\!\!D^{(1)}$. Both terms satisfy the property (\ref{gamma5}).  Let $v_i$ 
be the eigenvectors of $\not\!\!\!D^{(0)}$ and $v'_i$ the perturbed ones. The
 chirality of the perturbed eigenvectors to leading order in $\epsilon$ is 
\bea
{v'}^\dagger_i \gamma_5 v'_i = v^\dagger_i \gamma_5 v_i +  O(\epsilon)^2.
\label{mix} 
\eea
If $v_i$ is an eigenvector of $\not\!\!\!D^{(0)}$ with real eigenvalue, 
the first term on the right-hand side is of $O(1)$. Then a very large 
perturbation would be needed for 
 $v'_i$ to correspond to a complex eigenvalue of $\not\!\!\!D$ since, in this
case, the 
left-hand side of (\ref{mix}) vanishes. Similarly, for a complex eigenvalue of $\not\!\!D^{(0)}$ with eigenvector $v_i$ to become real, the left-hand side
of (\ref{mix}) should be of $O(1)$, while the first term on the right-hand side vanishes. A perturbation of $O(1)$ is needed.

The continuum index theorem implies that the number of zero modes with positive chirality minus the number of eigenvalues with negative chirality should be 
equal to the topological charge. 
As the lattice definition of the topological charge we will choose the 
geometrical definition of L\"uscher \cite{geo}, $Q^{geo}$. This charge 
is defined as the naive charge of a continuum gauge configuration obtained
by smoothly interpolating  the lattice configuration to the continuum. 
A continuum gauge field $a_{\mu}(x)$ can be constructed out of the link 
variables, which satisfies
\bea
U_{\mu}(s) = e^{i \;\int^{s+\hat{\mu}}_s dx \;a_{\mu}(x)},
\label{inter}
\eea
and which transforms covariantly under a lattice gauge transformation.
For Yang-Mills in four dimensions, $Q^{geo}$ is then defined as
\bea
Q^{geo} \equiv -\frac{1}{16\pi^2} \int d^4 x \tilde{f}_{\mu\nu} f_{\mu\nu},
\eea  
where $f_{\mu\nu}$ is the field strength of the field $a_\mu$.

Numerically, 
it has been found in all previous investigations (see for example \cite{itoh}) that the number of real 
eigenvalues of the Wilson-Dirac operator is a multiple of $2^d$ and their net
chirality vanishes. As we will see, this is the consequence of fermion 
doubling in non-trivial backgrounds.
On the other hand, the Wilson term is responsible for giving 
a large mass to the doublers, which translates in the fact that 
the $2^d$ sets of real eigenvalues cluster in $d+1$  
regions of the real axis. 
For smooth backgrounds, the real eigenvalues appear near the points $\lambda = 
2 r n_i/a$, where $r$ is the Wilson coupling, $a$ is the lattice spacing, $n_0=0$ for the physical modes, and $n_i$, for $i=1,...,2^d-1$, counts the number of momentum components that are equal to $\pi$ in the 
$i$th doubler corner of the Brillouin zone. 
We then define the physical region as the interval of the 
real axis $S_p = [0, r/a]$, where we expect to find the modes corresponding to 
the the physical fermion, and the doubler regions as $S_{d_i} = [(2 n_i -1) r/a, (2 n_i + 1 ) r/a]$. Notice that the regions corresponding to different
doublers with the same $n_i$ coincide.

 This pattern of the distribution of the real eigenvalues of $\not\!\!D$ is
 actually easy to prove for smooth
backgrounds (to my knowledge this has not been proved before).
Let us consider the lattice Wilson-Dirac operator (without bare mass):
\bea
\not\!\!D & \equiv & \frac{1}{2} \sum_{\mu} \gamma_{\mu} 
[  D^+_{\mu} + D_{\mu}^- ] - \frac{r}{2} \sum_{\mu} D_{\mu}^+ D^-_{\mu},
\label{Dlatt}
\eea
where the covariant are given by $D_{\mu}^+ \equiv \delta_{s's+\hat{\mu}} U_{\mu}(s) - \delta_{s's}$, $D_{\mu}^- \equiv \delta_{s's} - U^\dagger_{\mu}(s-\hat{\mu}) \delta_{s's-{\mu}}$. We also define the
 continuum Euclidean operator:
\bea
\not\!\!D^{c} \equiv \gamma_\mu (\partial_{\mu} + a_{\mu}),
\eea
where $a_{\mu}$ is the continuum field that satisfies eq. (\ref{inter}) and whose
topological charge is $Q^{geo}$. The continuum index theorem ensures that 
$\not\!\!\!D^{c}$ has zero modes
with a net chirality $Q^{geo}$. On the other hand, the lattice operator 
(\ref{Dlatt}) can be expanded (at small lattice momentum) in the lattice 
spacing, $a$,
\bea
\not\!\!D = \not\!\!D^{c} + a \;D_{\mu}^{c} D^{c}_{\mu} + ... 
\label{a1}
\eea
The terms of $O(a)$ are small, because we are considering smooth backgrounds and small lattice momentum.
 Then the zero modes of $\not\!\!D^{c}$ become small real eigenvalues of $\not\!\!D$. 
This is because the perturbation of the small $O(a)$ corrections can only 
move the zero modes 
along the real axis, but not make them complex, as we have explained before.

Considering the doubler fermions,  we must perform a similar expansion in $a$, but
around the appropiate corner of the Brillouin zone. This can be achieved by 
first performing a unitary transformation of the lattice operator and then 
expanding naively in $a$,
\bea
\not\!\!D^{(i)} \equiv V_{d_i}^\dagger \not\!\!D V_{d_i}= \frac{2 r n_i}{a} I + \not\!\!D^{c} + a\; D_\mu^{c} D_\mu^{c} + ..., 
\label{a2}
\eea
where
\bea
V_{d_i}(s) = \prod_{\nu} ( \delta_{K_i^\nu 0} + i \gamma_{\mu} \gamma_5 \delta_{K_i^\nu \pi} ) \;\exp(i\; K_i s),
\eea
 with $n_i=1,..,d$. $K_i$ is the $i$th doubler 
momentum (e.g. the lightest doublers in four dimensions correspond to $K_i = (\pi,0,0,0), 
(0,\pi,0,0), (0,0,\pi,0), (0,0,0,\pi)$ with $n_i =1$). It is easy to check 
that applying the $V_i$ rotation to the free lattice operator, we map 
the small momentum region into the region surrounding $K_i$. 
Again, for smooth backgrounds, the $O(a)$ terms are small, so the zero 
modes of $\not\!\!D^c$ become real modes of $\not\!\!D^{(i)}$ with 
eigenvalues $2 r n_i/a + O(a)$ (again the shift of the eigenvalues by the $O(a)$ effects is along the real axis).  On the other hand, applying $V^\dagger_i$ to the eigenvectors of $\not\!\!D^{(i)}$, we obtain eigenvectors of the 
$\not\!\!D$ with the same eigenvalues, which implies that $\not\!\!D$ 
also has real eigenvalues at $2 r n_i/a +O(a)$. This shows that for each
zero mode of $\not\!\!D^{c}$ in a smooth background, there are $2^d$ real modes
of $\not\!\!D$. Furthermore, if we define the lattice chirality of each eigenvector 
$v_i$ as
\bea
\chi_i \equiv sign(v^\dagger_i \gamma_5 v_i). 
\eea 
the net chirality in each  of the $2^d$ sets is $(-1)^{n_i} Q^{geo}$, since the $V_i$ for odd $n_i$ are chirality-flipping matrices. 
   
The lattice version of the index theorem then becomes,
\bea
N_R - N_L  = Q^{geo},
\label{lith}
\eea
 where $N_{R,L}$ are the real eigenvalues in $S_p$ with positive and 
negative chirality.

As we have shown, (\ref{lith}) is satisfied for smooth backgrounds, but not 
necessarily when rough gauge fields are considered. In general rough gauge 
configurations
are important even at small coupling, since they are responsible for 
renormalization. It is then not clear whether the two integer definitions
of topological charge, $Q^{geo}$ and $N_R-N_L$, and other quantities
constructed out of them, are equivalent up to $O(a)$ effects or if 
they differ even in the continuum limit. 

In two dimensions,  (\ref{lith}) will be exact at small enough coupling. The reason is that 
the effects of the higher dimensional operators of (\ref{a1}) and (\ref{a2}) are truly 
$O(a)$, because the integration over gauge fields in two dimensions can at most 
give logarithmically divergent contributions, which cannot compensate the 
naive power suppresion. This is of course due to the superrenormalizability 
of the theory. 
In four dimensions, this is not so obvious. The higher dimensional operators in (\ref{a1}) 
and (\ref{a2}), upon gauge field integration, will induce for example
a divergent renormalization of the fermion mass, which will 
shift all the zero modes of $\not\!\!\!D^c$ by a constant (in general it can be a 
different constant with alternating signs in $S_p$ and in $S_{d_i}$). Let's call the 
averages of the eigenvalues in $S_p$ and $S_{d_i}$, $m_c$ and $m_{d_i}$ 
respectively. For smooth backgrounds, we saw that
 $m_c = 0$ and $m_{d_i} = 2 r n_i /a$. Doubler decoupling requires 
that this hierarchy be
maintained under renormalization, in such a way that the physical mass
of the lightest doubler $\Delta_1 \equiv 
m_{d_1} - m_c \sim 1/a$. On the other hand, in order to ensure the 
separation between the two lightest sectors, the dispersion of the real eigenvalues 
around their means $m_c$ and $m_{d_1}$, that we call $\sigma_c$ and $\sigma_{d_1}$, should be much smaller than this. 
It has been conjectured \cite{bardeen}\cite{nara2} that this dispersion 
goes to zero in the continuum limit. Although this seems reasonable, 
we have not been able to find a rigorous proof for it. On the other hand, 
if it is true, this would seem to imply that the effect of the higher 
dimensional operators in (\ref{a1}) and (\ref{a2}) is, up to corrections vanishing in the continuum limit, a shift of the zero modes of $\not\!\!D^c$  
to the values $m_c$ and $m_{d_i}$. (If these higher dimensional operators
would induce non-vanishing effects other than the mass renormalization,
 this would translate generically in a dispersion of the real eigenvalues 
which would not vanish in the continuum limit.)  A
consequence of this is that the two charges $N_R-N_L$ and $Q^{geo}$ are
the same in the continuum limit (in particular if one has dislocations 
so does the other). 

Actually, it is quite possible that all the violations of the LITh at 
strong coupling are 
related to the fact that the gap between $S_p$ and $S_{d_i}$ closes.   
This gap is illustrated in Fig. 1, where we show the distribution of real eigenvalues
of the standard Wilson-Dirac operator resulting from a quenched 
simulation at $\beta=3.0, 1.0$ in a two-dimensional U(1) model. At the larger
$\beta$ value, the 
distribution clearly signals the expected regions of the real axis. 
However at $\beta=1.0$ the mixing  
between these regions is very large, and as a result
 assigning real modes to the physical or doubler sectors becomes ambiguous.  
This might be considered as an effect of the non-decoupling of the doublers 
in the measurement of topology. 
\begin{figure}[t]   
\vspace{0.1cm}
\centerline{\epsfig{figure=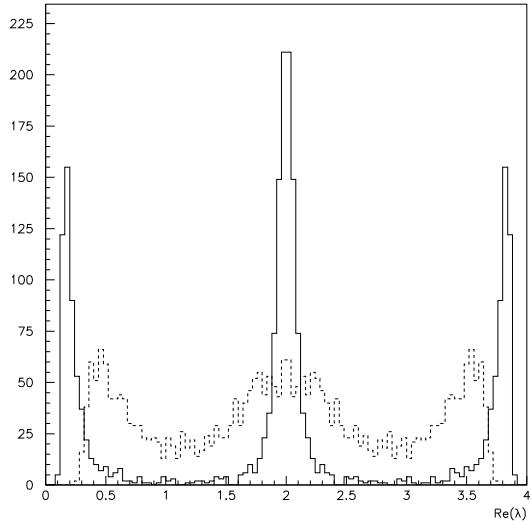,height=8cm,angle=0}}
\caption{Quenched distribution of the real eigenvalues values of the Wilson-Dirac operator in a two-dimensional $U(1)$ model at $\beta=3.0$ (solid) and $1.0$ (dashed),
at fixed physical volume.}
\end{figure}

If the values of $\sigma_{p,d_i}$ are truly $O(a)$, it is expected that 
they    
could be further reduced through standard improvement techniques to make 
them $O(a)^2$, in such a way that the gap between the different sectors 
be ensured for larger couplings. A first 
investigation of this has been presented in \cite{bardeen} and surprisingly 
a negative result has been found: the Clover term (which is the only operator of the appropriate dimensions and symmetries) does not seem to change $\sigma_p$.  In this letter, however, 
we propose a new improvement. We use ``improved'' in a 
somehow loose sense: the action is ``improved'' in the sense
that it has improved chiral properties. Since the $O(a)$ corrections 
of the standard Wilson action are chirality-breaking ones, the new
action has smaller $O(a)$ corrections. As we will see
in section 4, this action satisfies
\begin{eqnarray}
\not\!\!D^{impro} = \not\!\!D^c + O(\epsilon)
\label{impo}
\end{eqnarray}
at small momentum, with $\epsilon$ being a small parameter (at any gauge coupling) to be defined in section 4. This implies that $\sigma_p$ and $\sigma_{d_1}$, and  
also $|\Delta - 2 r/a|$ are controlled by $O(\epsilon)$ and not by the 
lattice spacing, ensuring in four dimensions a clean splitting between doubler 
sectors. Actually, (\ref{impo}) implies automatically that for small 
enough $\epsilon$, the (\ref{lith}) is satisfied for {\it arbitrarily} large couplings. The proof is identical to the one
 we gave for the standard action on 
smooth background gauge fields. Futhermore, one can also show that the 
chirality of the real modes is $\pm 1 + O(\epsilon)$, approaching closer
the continuum behaviour. 

Before discussing our results, it is worth pointing out that the overlap method to measure the 
topological charge \cite{nara} is also equivalent to the left-hand side of 
eq. (\ref{lith}). The charge in \cite{nara} 
is related to the number of level crossings of the 
Hamiltonian $H(\mu) \equiv \gamma_5 ( \not\!\!D - \mu )$. Clearly at the values
of $\mu$ at which $H(\mu)$ has a vanishing eigenvalue, the operator $\not\!\!D-\mu$ has a zero mode, with the same eigenvector, since this operator and 
$H(\mu)$ have a common kernel. On the other hand, the eigenvectors of $\not\!\!D_\mu$ and 
$\not\!\!D$ are the same for any $\mu$, so the zero modes of the first 
correspond to real eigenvalues of the second with eigenvalue $\mu$. Then 
the level crossings of $H(\mu)$ 
occur exactly at the values of $\mu$ that coincide with the real eigenvalues
of $\not\!\!D$. The charge is then defined as the number of positive chirality
crossings, minus the number of negative ones, which is exactly the 
left hand side of eq. (\ref{lith}). 
Also in this case, one has to define what the physical region
is and look at the crossings that occur only at $\mu < M$, where M is a 
large enough scale to separate the physical from the doubler sectors (
in \cite{nara} it is taken as $M=1/a$, for $r=1$, which coincides 
with $S_p$). Clearly, 
the previous reasoning is then also applicable to the overlap method. In 
particular the problem of the absence of a gap between the real modes
of the physical and doubler sectors will be the same.

\section{LITh with the standard action}

 We have considered a quenched two-dimensional compact U(1) model. The gauge action is the 
standard plaquette action:
\bea
{\cal S}_{g} \equiv  - \beta\; \sum_s {\mbox Re}\; U_{12}(s),
\label{sg}
\eea
with $U_{12}(s) \equiv U_1(s) U_2(s+\hat{1}) U_1^\dagger(s+\hat{2}) U_2^\dagger(s)$. In the unquenched case the violation of LITh 
can certainly not be worse, since the effect of the fermion determinant
will tend to suppress the measure of those configurations with non-trivial
topology. For a similar study of the standard Wilson action in the unquenched
case see, \cite{gatt}. 

In the quenched Schwinger model, the distribution of the geometrical topological charge can be computed analytically. We recall the definition of the 
geometrical charge for this model:
\bea
Q^{geo} \equiv \frac{1}{2\pi} \sum_{s} \theta_{12}(s) = \sum_s \log(U_{12}(s)), \;\;\;|\theta_{12}| < \pi. 
\eea
This quantity is an integer. The fraction of quenched configurations 
with a given geometrical charge $Q$ is then given by
\bea
\frac{Z_Q}{Z} = \frac{1}{2\pi} \int^{\pi}_{-\pi} d\alpha \;e^{-i \alpha Q}\; [ \sum_\nu \frac{I_{\nu}(\beta)}{I_0(\nu)} \frac{\sin(\pi \nu + \alpha/2)}{\pi\nu + \alpha/2}]^{L^2}
\label{zq}
\eea
where $Z = I_0(\beta)^{L^2}$ and $I_\nu(x)$ are the usual modified Bessel
functions and $L$ is the number of lattice sites in each direction. Of course, we have checked
that this distribution is correctly reproduced by our Montecarlo routine. 
\begin{figure}[t]   
\vspace{0.1cm}
\centerline{\epsfig{figure=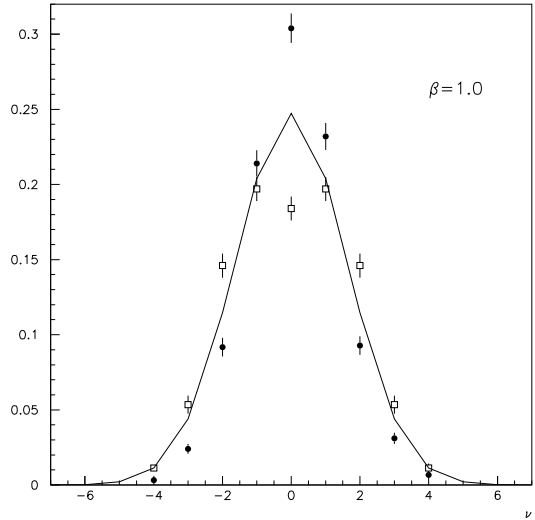,height=8cm,angle=0}}
\caption{Distribution of the values of $N_R-N_L$ (full circles) and $N^r/2^d$ (empty squares) in a lattice of $L=8$ at $\beta=1.0$, compared to the distribution of the geometrical charge (solid line).} 
\end{figure}
In Fig. 2, we compare this distribution to that of $N_R-N_L$ and 
$N^r/2^d$  calculated
numerically, for $\beta=1.0$ and $2\times 10^3$ configurations ($a\sim0.3$ in physical units, with 
respect to the mass gap in the continuum, i.e. $m = 1/\sqrt{\pi \beta}$).   
The quantity $N^r/2^d$, where $N^r$ is the total number of real eigenvalues, 
has been proposed by some authors as a possible
definition of topological charge \cite{gatt}. Clearly at this value of $\beta$, it is very distorted near zero.
  
On the other hand, the distribution of $N_R-N_L$ is also 
narrower than the geometrical one. 
This could be interpreted as due to the fact that the geometrical definition
can measure the topological charge of smaller lumps of charge, being then 
more sensitive to dislocations. Let us
suppose that the correct charge distribution was that obtained from the counting of real fermion modes. For each given configuration, the geometrical charge
will differ from the fermionic one by the presence of  one (to leading order, since a larger difference would be more rare) small lump of unit charge
of either sign. This means that the configuration will be counted with 
equal probability as having a geometrical charge smaller/larger in one 
unit with respect to the fermionic charge. Since the charge distribution is 
always peaked at zero and decreases for increasing charge, then more
configurations get shifted outwards than inwards, broadening it.

However, this broadening could also be explained in terms  of  
the mixing between the physical and doubler sectors of real eigenvalues.
Let us now suppose that the good distribution is the geometrical one;
then according to the LITh, the net chiral charge in $S_p$ is $Q^{geo}$. 
If $\sigma_p, \sigma_{d_1}$ are  
large (which will happen at strong coupling) 
the modes corresponding to the first doubler sector can get inside the 
region defined as $S_p$ or vice versa. Taking into account that the net 
chirality in the first doubler region is opposite to that in the physical region, it is easy to convince oneself, by considering several examples, that
the leading effect of this mixing is to narrow the fermionic charge distribution with respect to the geometrical one. 
If this 
is the mechanism by which the fermionic charge is not sensitive to very small
lumps of topological charge, it is clearly not very justified to say that the 
fermionic charge has less lattice artefacts. 

As we move towards smaller coupling, the three charge distributions
 become much closer, although there is always a systematic difference 
between $N^r/2^d$ and the other two.
 As we have argued, 
in the Schwinger model, when $\beta$ is increased, the chiral properties of action improve like $O(a/\beta)$.
\begin{table}
\begin{center}
\begin{tabular}{|c|c|c|c|c|c|}\hline
$ \beta$ & $L$ & $\langle\lambda\rangle_{p}$ & $\sigma_p$ & $\langle\lambda\rangle_{d_1}$ & $\sigma_{d_1}$ \\ \hline  
1 & 8  & 0.588 & 0.186& 1.596 & 0.281\\ 
1.6 & 10 & 0.484 & 0.192& 1.701 & 0.261\\
2.3 & 12 & 0.364 & 0.191& 1.818 & 0.216 \\ 
3.0 & 14 & 0.242 & 0.133& 1.902 & 0.144\\
4.0 & 16 & 0.151 & 0.057& 1.949 & 0.074\\
6.0 & 20 & 0.091 & 0.022& 1.981 & 0.017\\
\hline
\end{tabular}
\caption[]{Real Eigenvalue Distribution in lattice units 
for different $\beta$.}
\end{center}
\end{table}
 In Table 1, we present, for different 
$\beta$ values and fixed physical volume, the average and dispersion of the 
real eigenvalues (in lattice units) 
in the physical and first doubler regions. There  
$m_c \equiv \langle \lambda\rangle_p$, $\Delta-2 r =\langle\lambda\rangle_{d_1}$, $\sigma_{d_1}$ and $\sigma_p$ fall as $1/\beta$ for small enough coupling, confirming the expectation that they are $O(a)$ effects. Notice that the values of $\sigma_{p,d_1}$ are misleading at large couplings, because the corresponding distributions are far 
from being Gaussian  (see Fig. 1). 

 In Fig. 3 we present the probability to  
find the LITh being satisfied at fixed physical volume, for three values of 
this volume. Each 
point corresponds to 200 configurations (separated by 100 Monte Carlo sweeps for the smaller $\beta$ values and by 500 for the large $\beta$ values).
  The probability grows for smaller $1/\beta$ as shown 
in Fig. 3 up to a point $1/\beta_c$ where it saturates to $1$. The important 
point is that $\beta_c$ does not  seem to depend on the physical 
volume. The quantities $\sigma_p$ and $\langle\lambda\rangle_p$ also vary very little
 with  
the physical volume. An interpretation of Fig. 3 is  
that there is some critical 
splitting of the doubler sectors which ensures the validity of the LITh.
\begin{figure}[t]   
\vspace{0.1cm}
\centerline{\epsfig{figure=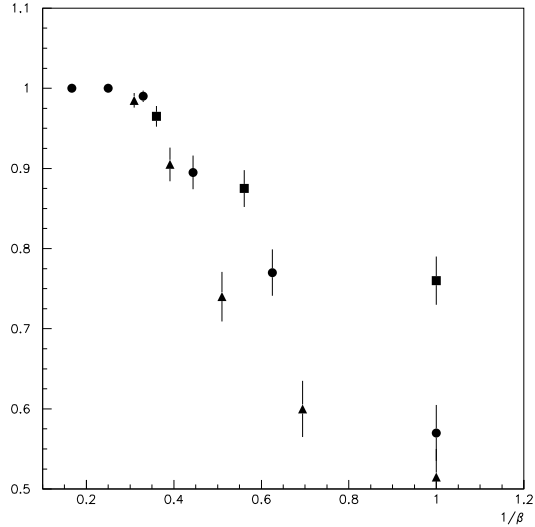,height=8cm,angle=0}}
\caption{Probability to find LITh satisfied as a function of $\beta^{-1}$ 
at fixed physical volume, for three values of the physical volume.} 
\label{qqtest}
\end{figure}

If the violations of LITh were related to small objects such as dislocations, one would expect a dependence of $\beta_c$ on the volume, since the entropy and thus the Boltzman weight of these artefacts grows with the volume at fixed $\beta$.
In this model there are no dangerous artefacts at weak coupling 
(e.g. dislocations), because 
there are no scale-invariant instanton solutions and, as the continuum
limit is approached, small lumps of topological charge are expected to be
strongly suppressed. So if we want to draw any conclusion about the 
effect of dislocations in four dimensions on the LITh from the results in this two-dimensional model, 
 the action should be improved while remaining at arbitrary 
strong coupling. This is possible with the action discussed in 
the next section.

\section{LITh for the Improved Action}
\label{sec:act}
From now on, the lattice spacing is denoted by $b$ to distinguish it from 
the lattice
spacing on the fermion lattice $f$ and from the lattice spacing for the 
standard action $a$. The $f$-lattice is some integer 
subdivision of the $b$-lattice. The $b$ sites are
denoted by $s$, while the sites on the finely-grained lattice are
$x$. The path integral is defined as 
\bea
Z[\eta]=\int {\cal D}U_{\mu} \;\; e^{-S_{g}[U]} \;\; det(\not\!\!D) \;\; e^{-\bar \eta {\not D}^{-1} \eta},
\eea
where $\eta$ are external fermion sources. The gauge action
 is the standard plaquette action of eq. (\ref{sg}) and the measure is the 
standard one in a lattice of spacing $b$. The only difference is in the
Wilson-Dirac operator, which is defined on the $f$-lattice as follows:
\bea
\not\!\!D & \equiv & \frac{1}{2} \sum_{\mu} \gamma_{\mu} 
[  D^+_{\mu} + D_{\mu}^- ] 
- \frac{r}{2} \sum^{2}_{\mu=1} D_{\mu}^+ D^-_{\mu},
\label{u1}
\eea
where the covariant and normal derivatives are given by $D_{\mu}^+ \Psi(x) = u_{\mu}(x) \Psi(x+\hat{\mu}) - \Psi(x)$, $D_{\mu}^- \Psi(x) = \Psi(x) - u^\dagger_{\mu}(x-\hat{\mu}) \Psi(x-\hat{\mu})$. The $u_{\mu}(x)$ link variables are interpolations of the real dynamical 
fields $U_\mu(s)$ \cite{us2}. The reader is referred to \cite{us2} for details
on the method to construct the interpolation for non-Abelian theories. 
In this $U(1)$ model the interpolation is particularly easy. Defining the sites on the $f$-lattice as $x = s b + t_1 f \hat{1} + t_2 f \hat{2}$, where $0 \leq t_1, t_2 \leq 1$, the interpolation that we use for compact $U(1)$ in two dimensions is given by,
\bea
u_1(t_1,t_2) &=& \omega^\dagger(t_1,t_2)\; [U^\dagger_{12}(s)]^{t_2\; \frac{f}{b}} \;\omega(t_1+\frac{f}{b},t_2)\; \nonumber\\
u_2(t_1,t_2) &=& \omega^\dagger(t_1,t_2)\;\omega(t_1,t_2+\frac{f}{b})\; \nonumber\\
\label{interqed}
\eea
where the $\omega$ fields are defined as
\bea
\omega(0, t_2) &=& U_2(s)^{t_2}, \nonumber \\
\omega(1, t_2) &=& U_1(s) \;(U_2(s+\hat{1}))^{t_2},\nonumber \\
\omega(t_1, t_2) &=& (U_1(s)^{t_1})^{1-t_2}\; [U_{12}(s)^{t_1}\; U_2(s)\; U_{1}(s + \hat{2})^{t_1}]^{t_2}. 
\label{interqed2}
\eea
It is trivial to show that this interpolation is gauge covariant, rotationally
invariant up to a gauge transformation, and strictly local\footnote{Notice that
this interpolation is not the same one proposed in \cite{us2}. The reason is that (\ref{interqed}) is not appropiate for chiral gauge theories, because the 
interpolated gauge field is not bounded.}.

The continuum limit of this theory is defined as
\begin{eqnarray}
\xi^{-1} b \rightarrow 0 \;\;\;\;\;\; 
\end{eqnarray}
for $f/b$ fixed. 
The real cutoff is $b$.  We can actually construct the improved 
fermion operator
in terms of $U_{\mu}$, taking any number for $f/b$, without ever referring to 
the lattice $f$. 

From the power counting arguments of \cite{us1}, the breaking of chiral
 symmetry is suppressed by at least a power of the ratio $f/b$ to all 
orders in the gauge coupling. In the appendix, we compute the one-loop 
correction to the quark mass renormalization in QED in four dimensions, to leading
order in $f/b$, and show this suppression by explicit calculation.
It is not hard to understand this suppression along the lines of section 2. 
The continuum field $a_{\mu}$ can easily be chosen to also satisfy
$u_{\mu}= \exp( i \int a_\mu )$. The operator (\ref{u1}) can be expanded in $f$
to give
\bea
\not\!\!D = \not\!\!D^c + f \; D^c_\mu D^c_\mu + ... = \not\!\!D^c + O(f/b)
\label{impo2}
\eea
The higher-dimensional operators are truly of $O(f)$, in contrast 
with the standard action, because the roughness of the continuum field $a_\mu$
is at most of $O(1/b)$. Similarly, close to the doubler momenta, 
\bea
\not\!\!D^{(i)} = \frac{2 r n_i}{f} + \not\!\!D^c + f \; D^c_\mu D^c_\mu + ...
= \frac{2 r n_i}{f} + \not\!\!D^c + O(f/b). 
\eea
For finite $f/b$, (\ref{impo2}) implies that the LITh is satisfied for 
arbitrarily large coupling, by the arguments of section 2. Furthermore, the
 real modes in $S_p$ have magnitudes of $O(f/b)$, without any tuning to the 
chiral limit and their quiralities are $\pm 1$ up to $O(f/b)$ corrections. 

All these expectations are confirmed numerically. In Fig. 4 we show the
 dispersion 
of the real eigenvalues in $S_p$, for a simple fine-graining factor of $1/2$. 
The values have been shifted to an average value of zero. 
\begin{figure}[t]   
\vspace{0.1cm}
\centerline{\epsfig{figure=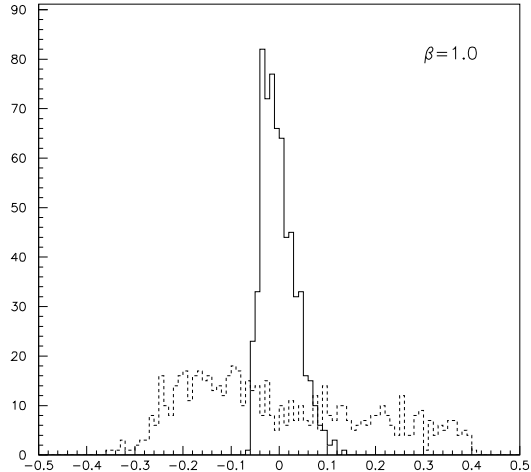,height=8cm,angle=0}}
\caption{Distribution of real eigenvalues (in $f$ units) for $f/b=1/2$ (full line), $f/b=1$ 
(dashed line) for a lattice size of $L_b=8$ and for $\beta=1.0$.} 
\label{real}
\end{figure}
The improvement in the magnitude and dispersion of the real eigenvalues 
is roughly $(f/b)^2$ (in $f$ units), as expected. This is in contrast with the results of 
\cite{bardeen}, concerning the Clover improved action for this model. 

On the other hand, the improvement on the LITh is much more 
dramatic. In Fig.~5, 
we show the probability of finding the LITh, i.e. $Q^{geo}= N_R-N_L$, as
a function of $\beta$ for a $b$-lattice $L_b = 8$ and a
fine-graining factor of $1$ (i.e. $f=b=a$) and $1/2$ (i.e. $f=b/2$). The 
number of uncorrelated configurations is $500$ (separated by 500 sweeps) 
in each case. No violations of the LITh
have been observed for the improved action for arbitrary rough fields. 
\begin{figure}[t]   
\vspace{0.1cm}
\centerline{\epsfig{figure=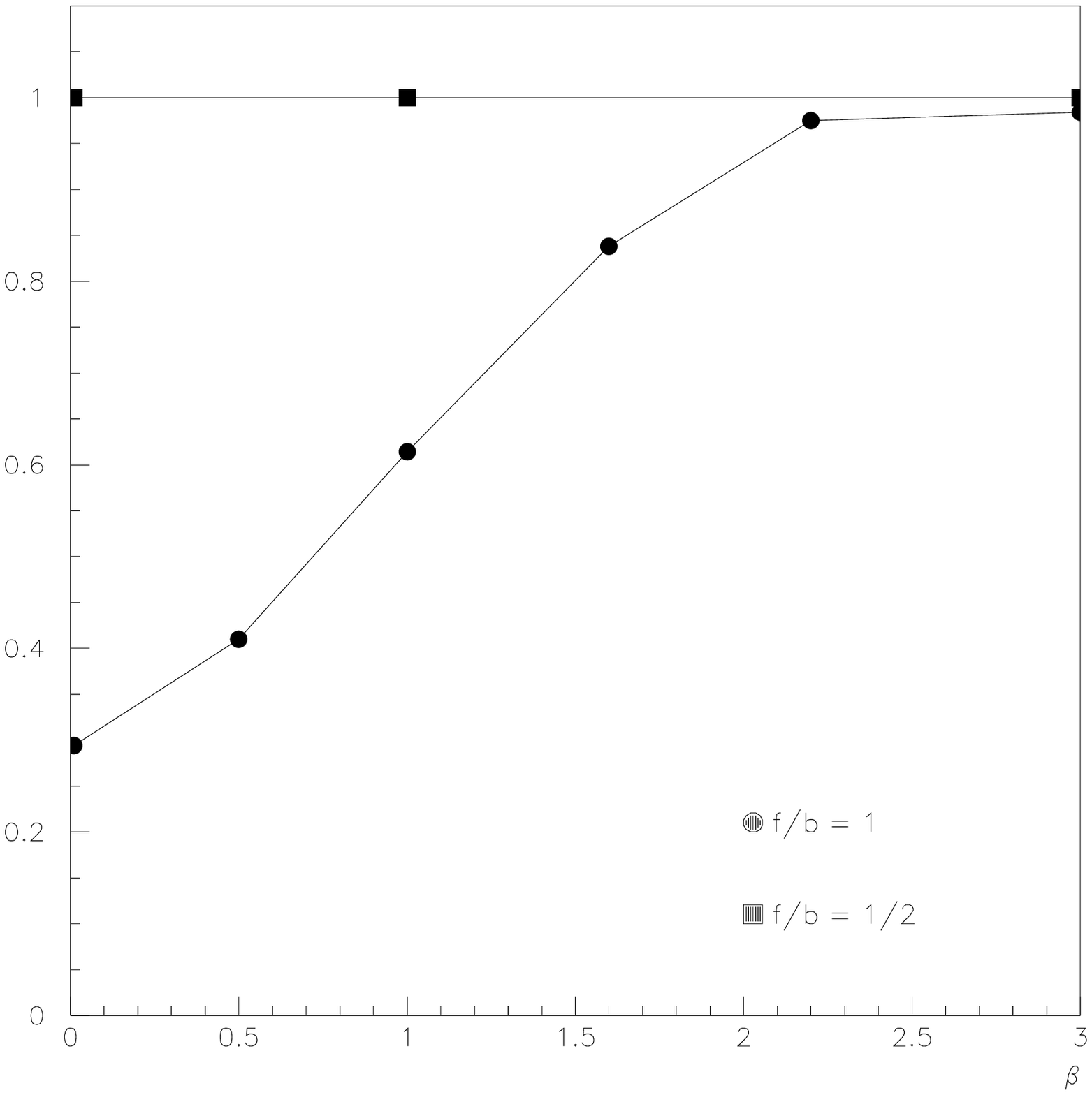,height=8cm,angle=0}}
\caption{Probability of finding $Q^{geo} = N_R - N_L$ in a lattice of $L_b=8$ as
a function of $\beta$ for $f/b=1$ (circles) and  $1/2$ (squares).} 
\label{prob}
\end{figure}
Also for fixed $\beta$, the probability remains $1$ at larger physical volumes.
On the other hand, it is worth pointing out that, even 
for the improved action, 
the charge defined as $N^r/2^d$ does not coincide with the geometrical one. 
For example, the probability of their agreement is $0.93$ at $\beta=0.01$, 
improving slowly for larger $\beta$.
 
The fact that the LITh  is satisfied for the improved action at arbitrarily
strong coupling indicates that, as expected, similar results should hold in four dimensions. If 
this is so, there is no reason to believe 
that the fermion method to measure topological charge has less artefacts than
the geometrical one. The problem of dislocations is related to the gauge action, which is very poor at measuring the action of small sized objects carrying
topological charge, and the sensible
way to get rid of this problem is to improve the gauge action \cite{marga,hasen,us3}.  
On the other hand, measuring topology through the counting of the small real 
eigenvalues (taking into account their chiralities) might be 
computationally more efficient. A study of this issue
will be presented elsewhere. 

\section{Discussion and Conclusions}
\label{sec:wis}

It is well known that the quenched approximation has problems near the 
chiral limit. In the so-called exceptional configurations, it 
is very hard or impossible to invert the fermion matrix. Recently, a cure has been proposed for
this \cite{bardeen}. Exceptional configurations 
occur whenever there are real eigenvalues that get close to zero. 
As is clear from Fig. 4, this is bound to happen if the subtracted
fermion mass is smaller than $\sigma_p$. 
The proposal of \cite{bardeen} is that whenever an exceptional 
configuration is found, the real eigenvalues must be shifted by the appropriate
amount to be exactly zero at the chiral point, which implies that 
the fermion matrix is invertible arbitrarily close to this point.
As already stated in \cite{bardeen}, this is a non-local procedure, which might 
introduce unphysical effects. It is unclear how one could be sure
about the safety of this method, even if it seems to give good results
for a given observable. 

In the light of  our results, 
it is important to stress that exceptional configurations
 are simply topologically non-trivial configurations. Clearly the quenched approximation should not be 
well-behaved in non-trivial configurations, since a big suppression
in the Boltzman weight of such configurations comes from the fermion 
determinant, which is being neglected. A much more justified procedure
would be to ignore topologically non-trivial configurations in 
this approximation. Alternatively one should not
 approach the chiral limit closer than $\sigma_p$ ($O(f/b)^2$ for 
the improved action or supposedly $O(a)$ for the standard action), i.e. work with a quark mass $\geq \sigma_p$. This is a sensible thing to do even for the 
standard action if,  as
argued in \cite{bardeen} \cite{nara2}, the dispersion of the real 
eigenvalues is a lattice artifact
that vanishes in the continuum limit. In other words, the physical 
fermion mass cannot be determined with a better accuracy than $O(a)$ \cite{nara2}, 
so working with a physical mass of order $\sigma_p = O(a)$ is not
surprisingly the safe choice. 

The lesson is that the physical mass should then not be reduced without improvement. 
At least in principle $f/b$ can 
be taken to be as small as needed \footnote{Of course lowering
$f/b$ is hard computationally, but it should be possible to parametrize 
the action for arbitrarily small $f/b$ in terms of a $b$ lattice
fermionic action with higher dimensional operators with coveniently tuned
 coefficients.}. Our improved action then has the same effect 
as the ``modified'' quenched approximation, i.e. one can get closer to 
the chiral point without encountering exceptional configurations, except 
that it does it in a local way. For another possible solution to this 
problem, see \cite{jansen}.

For full dynamical fermions, the problem of exceptional configurations 
is not so important, because they have a very small Boltzman weight. 
Of course this means that they are less probable, but can and will appear. 
On the other hand, it is clear that configurations with real eigenvalues
are a serious problem for unquenched simulations with an odd number of 
flavours. 
Figure 4 shows that if the renormalized mass is smaller than $O(\sigma_p)$, we will often find configurations
with a real negative eigenvalue (notice that complex eigenvalues come 
in complex conjugate pairs, so their contribution to the determinant 
is always positive), which makes the determinant negative;  
the Euclidean measure is thus
 not positive-definite and Montecarlo methods will 
fail. Again the obvious solution is to choose a quark mass of
$O(\sigma_p)$ or treat the negative eigenvalues as in \cite{jansen}.

To conclude, 
we have presented evidence for the exact validity of the index theorem on the
lattice if the zero modes are identified with the exactly real eigenvalues
of an improved Wilson-Dirac operator in the physical region (i.e. $\lambda < r/a$) and if the gauge topology is measured with the geometrical definition 
of ref. \cite{geo}. We have argued that the violations of this
``theorem'' that are observed for rough configurations are related to the 
large dispersion of the real eigenvalues: the different doubler regions cannot
be cleanly separated in the real axis and the counting of the physical 
chiral modes gets contaminated by the doublers. The effect of
 improving the fermionic action is to ensure a finite gap  
between the doubler sectors, which then ensures the exact validity of the 
``index theorem'' for arbitrarily large coupling. 
These results strongly suggest that fermionic methods of measuring 
an integer-valued 
topological charge using the Wilson operator are equivalent to the 
geometrical definition.

A few days ago two new preprints appeared which study
 the index theorem for two other types of improvement: perfect actions \cite{hasen2} and Clover improvement in the second \cite{simma}. Actually our improved fermionic action in the limit $f/b\rightarrow 0$, combined with the improved 
gauge action of \cite{us3}, shares many properties with perfect actions.  Not
surprisingly the index theorem becomes exact in both cases at arbitrarily 
large gauge coupling.

\section*{Acknowledgements} 
My special thanks to K. Jansen and A. Vladikas for their very 
critical reading of the first draft of this manuscript.  I also thank them 
as well as A. Donini, M. Guagnelli and M. Testa for very useful discussions. 

\section{Appendix A} 
In this appendix, we sketch the one-loop calculation of the mass 
renormalization in QED in four dimensions with the two-cutoff action. In perturbation 
theory, the interpolation of \cite{us2} simplifies greatly, and 
an analytic formula can be obtained, which corresponds to a linear 
interpolation\footnote{Notice that in lattice perturbation theory there
are no problems with the windings discussed in \cite{us2}.}. 
The Fourier transform of the interpolated in field in four dimensions is, 
\begin{eqnarray}
a_{\mu}(q) = \frac{1}{n^4} A_{\mu}(\bar q) \frac{e^{i q_{\mu} n} - 1}{e^{i q_{\mu}} - 1}
\prod_{\alpha\neq \mu} \frac{1}{n} \frac{e^{i q_{\alpha} n} 2 (\cos z_{\alpha}-1)}{(e^{i q_{\alpha}} - 1)^2},
\label{lin}
\end{eqnarray}
where $q_{\mu} = \bar q_{\mu} + \frac{2 \pi m_{\mu}}{n}$, $|\bar q_{\mu}| < \frac{\pi}{n}$ and the integers 
$m_{\mu} = -n/2,...,n/2$. $z_{\mu} \equiv q_\mu n = {\bar q}_{\mu} n$ Notice that at low momentum, $a_{\mu}(q) 
\rightarrow A_{\mu}(\bar q)$, as expected from the locality of the 
interpolation.

In order to derive perturbation theory, we need to go to momentum space. Except for the gauge-field propagator, 
all the lattice Feynman rules are 
the same as for standard Wilson fermions on the $f$ lattice. We 
will use $f=1$ units to simplify the formulae. Then we should extract
 the leading dependence on $n\equiv b/f$, which is the factor of fine-graining. 

 From (\ref{lin}), the propagator of the 
interpolated gauge field can be obtained in terms of the gauge propagator
on the $b$ lattice in the Feynman gauge: 
\begin{eqnarray}
\langle a_{\mu}(q) a_{\nu}(-q) \rangle = \delta_{\mu\nu}
\frac{n^2}{2 \sum_{\rho} \sin(\bar{z}_{\rho}/2)^2} \frac{1}{n^8} \frac{1 -  \cos z_{\mu}}{1- \cos q_{\mu} } \prod_{\alpha\neq \mu} 
\frac{1}{n^2} (\frac{1-\cos z_{\alpha} }{1 - \cos q_{\alpha}})^2.
\end{eqnarray}

Now, we want to compute the linear divergence in the fermion self-energy. 
The two contributing diagrams are those shown in Fig. 5.
\begin{figure}[t]   
\vspace{0.1cm}
\centerline{\psfig{figure=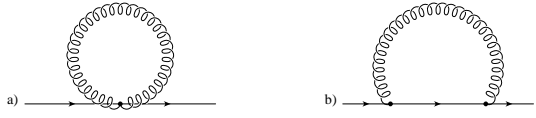,width=7cm}}
\caption{Diagrams contributing to the fermion self-energy at one loop.} 
\label{self}
\end{figure}
 In order to get the linear
divergence, the external fermion momenta can be set to zero. The tadpole contribution is given by
\bea
\Sigma^{a}(0) = - g^2 \frac{1}{n^4} \int_{BZ} \frac{d^4 z}{(2 \pi)^4} 
\frac{n^2}{2 \sum_{\rho} \sin(z_{\rho}/2)^2} \sum_{\mu} \frac{1}{n^8} \sum_{m_{\mu}=-n/2}^{n/2}  
\frac{1 -  \cos z_{\mu}}{1- \cos q_{\mu} } \nonumber\\
\prod_{\alpha\neq \mu} \sum_{m_{\alpha}=-n/2}^{n/2}  \frac{1}{n^2} (\frac{1-\cos z_{\alpha} }{1 - \cos q_{\alpha}})^2.
\eea
The integer sums can be analytically computed, using
\bea
\sum_{m_{\mu}=-n/2}^{n/2} \frac{1}{\sin(\bar q_{\mu}/2 + \pi m_{\mu}/n)^2} = 
\frac{2\; n^2}{1-\cos z_{\mu}}, \nonumber\\
\sum_{m_{\mu}=-n/2}^{n/2} \frac{1}{\sin(\bar q_{\mu}/2 + \pi m_{\mu}/n)^4} = 
n^4 \frac{4}{3} \frac{2 + \cos z_{\mu}}{(1-\cos z_{\mu})^2}.
\label{sums}
\eea
Finally, we get
\bea
\Sigma^{a}(0) = - g^2 \frac{1}{27\; n^2} \int_{BZ} \frac{d^4 z}{(2 \pi)^4} 
\sum_{\mu} \frac{\prod_{\alpha\neq\mu} ( 2 + \cos z_{\alpha})}{2 \sum_{\rho} \sin(z_{\rho}/2)^2} + O(\frac{1}{n^4}). 
\eea
The remaining integral is a finite number, which can be computed numerically. 
 The leading dependence on $n$ is thus the expected one.

The contribution from the second diagram is a little harder to obtain, 
because the fermion propagator enters the integer sums;  however, the 
leading dependence on $n$ can also be obtained analytically. The diagram 
gives
\bea
\Sigma^{b}(0) = - g^2 \frac{1}{n^4} \int_{BZ} \frac{d^4 z}{(2 \pi)^4} 
\frac{n^2}{2 \sum_{\rho} \sin(z_{\rho}/2)^2} \sum_{\mu} \frac{1}{n^8} \sum_{m_{\mu}=-n/2}^{n/2}  
\frac{1 -  \cos z_{\mu}}{1- \cos q_{\mu}}\nonumber\\
\prod_{\alpha\neq \mu} ( \sum_{m_{\alpha}=-n/2}^{n/2}  \frac{1}{n^2} (\frac{1-\cos z_{\alpha}}{1 - \cos q_{\alpha}})^2 )\;\;\frac{\cos q_{\mu} M(q) - \sin^2 q_{\mu}}{s(q) + M(q)^2},
\eea
with $M(q) = \sum_{\rho} 1 -\cos q_{\rho}$ and $s(q)= \sum_\rho (\sin q_\rho)^2$. The integer sum over $m_{\mu}$ can be derived from eqs. (\ref{sums}) after 
some manipulations:
\bea
\sum_{m_{\mu}} \frac{\cos q_{\mu}\; M(q) - \sin^2 q_{\mu}}{(1-\cos q_{\mu})\; (s(q) + M(q)^2)} = n^2 \frac{1}{1-\cos z_{\mu}} \frac{M'(q)}{s'(q) + M'(q)^2} + O(n)
\eea
where $s'(q) = \sum_{\rho\neq \mu} (\sin q_\rho)^2$ and $M'(q) = \sum_{\rho\neq\mu} 1 - \cos q_\rho$. The next sum required is
\bea
\sum_{m_\nu} \frac{M'(q)}{s'(q) + M'(q)^2} \frac{1}{(1-\cos q_{\nu})^2}
= \frac{n^4}{3} \frac{2 + \cos z_{\nu}}{(1-\cos z_{\nu})^2} \frac{M''(q)}{s''(q) + M''(q)^2} + O(n),
\label{sum}
\eea
with $s''(q)= \sum_{\rho \neq \mu,\nu} (\sin q_\rho)^2$ and similarly for 
$M''(q)$. 
The last two sums are also of the form (\ref{sum}) and the final result
is
\bea
\Sigma^{b}(0) = - g^2 \frac{1}{54 n^2} \int_{BZ} \frac{d^4 z}{(2 \pi)^4}
\frac{1}{2 \sum_{\rho} \sin^2(z_{\rho}/2)} \sum_\mu \; \prod_{\alpha\neq \mu} (2+ \cos z_\alpha ) + O(\frac{1}{n^3}).
\eea
Again the remaining integral is a finite number, so the $n$ dependence
is as expected.

\end{document}